\documentclass[rnote,oldversion,longauth]{aa} 
\usepackage{natbib,graphicx}
\bibpunct{(}{)}{;}{a}{}{,}

\begin{document}

\title{MAGIC Observations of PG\,1553+113 during a \\ Multiwavelength
Campaign in July 2006}


%
\author{
 J.~Albert\inst{1} \and 
 E.~Aliu\inst{2} \and 
 H.~Anderhub\inst{3} \and 
 P.~Antoranz\inst{4} \and 
 C.~Baixeras\inst{5} \and 
 J.~A.~Barrio\inst{4} \and 
 H.~Bartko\inst{6} \and 
 D.~Bastieri\inst{7} \and 
 J.~K.~Becker\inst{8}\and
 W.~Bednarek\inst{9}
% ...
}
 \institute {Universit\"at W\"urzburg, D-97074 W\"urzburg, Germany 
 \and IFAEnergies, Edifici Cn., E-08193 Bellaterra, Spain          
 \and ETH Z\"urich, CH-8093 H\"onggerberg, Switzerland             
 \and Universidad Complutense, E-28040 Madrid, Spain               
 \and Universitat Aut\`onoma de Barcelona, E-08193 Bellaterra, Spain 
 \and Max-Planck-Institut f\"ur Physik, D-80805 M\"unchen, Germany 
 \and Universit\`a di Padova and INFN, I-35131 Padova, Italy       
 \and Universit\"at Dortmund, D-44227 Dortmund, Germany            
 \and University of \L \'od\'z, PL-90236 Lodz, Poland              
% ...
}
 

\abstract{The active galactic nucleus PG\,1553+113 was observed by
the MAGIC telescope in July 2006 during a multiwavelength campaign, in
which telescopes in the optical, X-ray, and very high energies
participated. Although the MAGIC data were affected by strong
atmospheric absorption (calima), they were analyzed after applying a
correction. In 8.5~hours, a signal was detected with a significance of
5.0\,$\sigma$. The integral flux above 150\,GeV was
$(2.6\pm0.9)\cdot10^{-7}~{\rm ph\,s^{-1}m^{-2}}$. By fitting the
differential energy spectrum with a power law, a spectral index
of $-4.1\pm0.3$ was obtained.}

\titlerunning{MAGIC Observations of PG\,1553+113 during a MWL Campaign}

\authorrunning{J.~Albert et al.}

\date{Received 13.11.2007 / Accepted 16.10.2008 }

\offprints{Daniela Dorner, \email{daniela.dorner@unige.ch}}


\maketitle

\section{Introduction}

The Major Atmospheric Gamma-ray Imaging Cherenkov (MAGIC) telescope,
located on the Canary Island of La Palma at 2200\,m~a.s.l., is capable
of extending very high energy observations to energies previously
unreachable and detecting new sources at energies down to 50\,GeV. 

One of these sources is the BL Lac type object PG\,1553+113, observed
for the first time in this energy range in April and May 2005 by the
MAGIC telescope and the High Energy Stereoscopic System (H.E.S.S.).
From these observations a faint signal was measured, and the detection
was confirmed by further observations \citep{hess1553,magic1553}. In
the subsequent years, additional VHE data were taken. From the combined
data sets, strong signals were found: for H.E.S.S.\ at 10.2\,$\sigma$
significance \citep{hess1553-2} and for MAGIC at 15.0\,$\sigma$
significance \citep{phd}. In addition, the VHE measurements allowed the
unknown redshift of the source to be constrained. Until now the
redshift of PG\,1553+113 has been unknown, since neither emission or
absorption lines could be found, nor the host galaxy resolved. Several
lower limits were determined
\citep{carangelo,Sbarufatti2005,Sbarufatti2006,Scarpa2000,Treves2007}.
With the MAGIC and H.E.S.S.\ measurements, upper limits could be
determined \citep{hess1553,mazin,phd}.

To study the spectral energy distribution of a source, simultaneous
data from different wavelengths are mandatory. Therefore, a
multi\-wave\-length (MWL) campaign was performed in July 2006 to
observe PG\,1553+113. This paper concentrates on the data taken by
MAGIC during this MWL campaign.

\section{Observations and Data Quality}

Between April 2005 and April 2007, MAGIC observed PG\,1553+113 for a
total of 78~hours. Part of these data were acquired during a MWL
campaign in 2006 July with the H.E.S.S.\ array of IACTs, the X-ray
satellite Suzaku and the optical telescope KVA. Suzaku observed the
source between 24 July, 14:26~UTC and 25 July, 19:17~UTC, and H.E.S.S.
between 22 July and 27 July. From the KVA, data between 21 July and 27
July are available. 

The MAGIC telescope observed PG\,1553+113 between 14 July and 27 July
for 9.5~hours at zenith distances between 18$^\circ$ and 35$^\circ$.
The data were acquired in wobble mode, where the source was tracked
with an offset of $\pm0.4^\circ$ from the center of the camera,
which enabled simultaneous measurement of the source and the
background. 

One hour of data was excluded due to technical problems. The quality of
the entire data set acquired during the MWL campaign was affectd by
calima, i.e.\ sand-dust from the Sahara in the atmosphere. For the
affected data, the nightly values of atmospheric absorption ranged
between 5\% and 40\%. To account for the absorption of the Cherenkov
light, correction factors were calculated and applied to the data (see
Sect.~\ref{corr}).

\section{Analysis}

The data were processed by an automated analysis pipeline
\citep{autom,mars} at the data center in W\"urzburg. The analysis
includes an absolute calibration with muons \citep{muoncal}, an
absolute mispointing correction \citep{starg}, and it uses the arrival
time information of the pulses of neighboring pixels for noise
subtraction and background suppression. 

In determining the background, three OFF regions were used, providing a
scale factor of 1/3 for the background measurement. 

To suppress the background, a dynamical cut in Area
(Area=$\pi\cdot$WIDTH$\cdot$LENGTH) versus SIZE and a cut in
$\vartheta$ were applied. More details of the cuts can be found in
\citet{cuts}, and the aforementioned image parameters are described by
\citet{hillas}. To account for the steeper spectrum of PG\,1553+113,
the here presented analysis applied a cut in Area that was less
restrictive at lower energies compared to the standard cut used by the
automated analysis, which has been optimized for Crab Nebula data
over several years. 


In generating the spectrum, the cut in Area was selected to ensure 
that more than 90\% of the simulated gammas survived. To study the
dependency of the spectral shape on the cut efficiency, a different cut
in Area with cut efficiencies of between 50\% and 95\% for the entire
energy range was applied.

\section{Correction of the Effect of Calima}
\label{corr}

Calima, also known as Saharan Air Layer (SAL), is a layer in the
atmosphere that transports sand-dust from the Sahara in a westerly
direction over the Atlantic Ocean. The SAL is usually situated between
1.5\,km and 5.5\,km~a.s.l.\ \citep{sal}. Since the Canary Islands are
close to the North African coast, the MAGIC observations are probably
affected by additional light absorption in the atmosphere when calima
occurs.

From extinction measurements of the Carlsberg Meridian Telescope 
\citep{cmt,cmtweb}, which are available for each night, the loss of
light due to calima was calculated. To correct the data for absorption
by the atmosphere, the calibration factors were adapted for each night
separately and the data were reprocessed. More details of the method
are provided by \citet{calima}.

\section{Results}

\begin{figure}[ht]
\begin{center}
 \resizebox{\hsize}{!}{\includegraphics{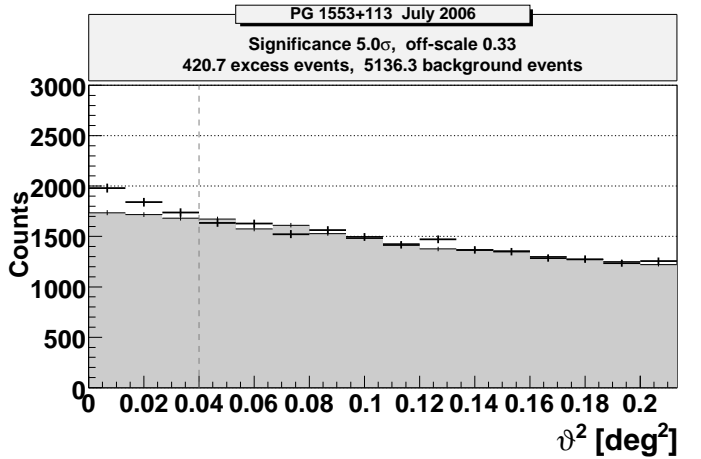}}
\caption{Distributions of $\vartheta^2$ for ON-source events (black
crosses) and normalized OFF-source events (gray shaded area) from
8.5~hours of data. The dashed line represents the cut applied in
$\vartheta^2$. \label{thetasq}} 

\end{center} 
\end{figure}

The 8.5~hours of data from PG\,1553+113 provided a signal with a
significance of 5.0\,$\sigma$ according to \citet{lima}. The
$\vartheta^2$ distributions for the ON- and normalized OFF-source
measurement are shown in Fig.~\ref{thetasq}.

\begin{figure}
\begin{center}
 \resizebox{\hsize}{!}{\includegraphics{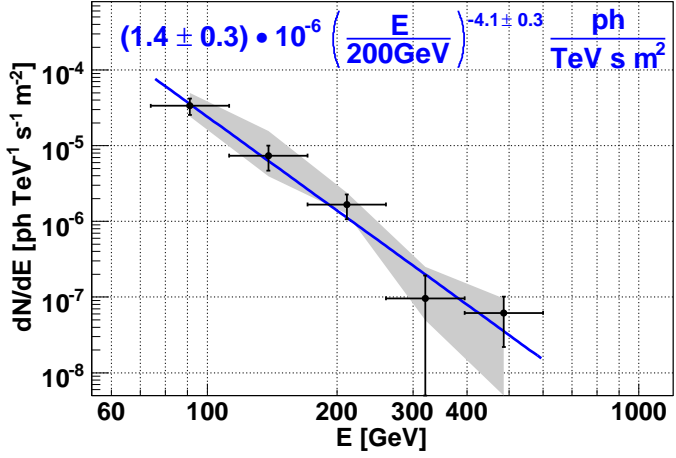}}
\caption{Differential energy spectrum of PG\,1553+113. The horizontal
error bars indicate the width of the energy bins. The vertical error
bars illustrate the statistical errors. The gray band corresponds to a set of
spectra obtained with different cut efficiencies and simulated spectra.
\label{spectrum}} 
\end{center}  
\end{figure}

The differential spectrum measured by MAGIC is shown in
Fig.~\ref{spectrum}. In this plot, a set of spectra obtained with
different cut efficiencies and different simulated spectra is indicated
by a gray band. The data points of the spectrum are given in
Table~\ref{tablespectrum} including the statistical errors. The
systematic errors of the analysis are discussed in \citet{crab}.

Fitting the differential spectrum with a power law yields a flux of
$(1.4\pm0.3)\cdot10^{-6}~{\rm ph\,TeV^{-1}s^{-1}m^{-2}}$ at
200\,GeV and a spectral index of $-4.1\pm0.3$. This result is
consistent with the data (fit probability 45\%) and in good agreement
with previous measurements in 2005 and 2006 \citep{hess1553,magic1553}.
Within the errors, the simultaneous measurements of the H.E.S.S.
telescopes in the energy range above 225\,GeV is in agreement with the
MAGIC results, although the fit of the differential spectrum yields a
spectral index of $-5.0\pm0.7$ \citep{hess1553-2}. 
The integral flux above 150\,GeV obtained from this analysis is
$(2.6\pm0.9)\cdot10^{-7}~{\rm ph\,s^{-1}m^{-2}}$.

\begin{table}
\caption{Flux for the spectral points in Fig.~\ref{spectrum}
including the statistical errors.}
\label{tablespectrum}
\centering
\begin{tabular}{c c c}
\hline\hline
 E         &  F                                 & Statistical Error \\
 $[$GeV$]$ & [ph\,${\rm TeV^{-1}s^{-1}m^{-2}}$] & [ph\,${\rm TeV^{-1}s^{-1}m^{-2}}$]\\
\hline
91      & 3.36$\cdot10^{-5}$ & 8.11$\cdot10^{-6}$ \\
139     & 7.37$\cdot10^{-6}$ & 2.68$\cdot10^{-6}$ \\
211     & 1.67$\cdot10^{-6}$ & 6.04$\cdot10^{-7}$ \\
320     & 9.61$\cdot10^{-8}$ & 9.61$\cdot10^{-8}$ \\
486     & 6.18$\cdot10^{-8}$ & 4.00$\cdot10^{-8}$ \\
\hline
\end{tabular}
\end{table}

\begin{figure}
\begin{center}
 \resizebox{\hsize}{!}{\includegraphics{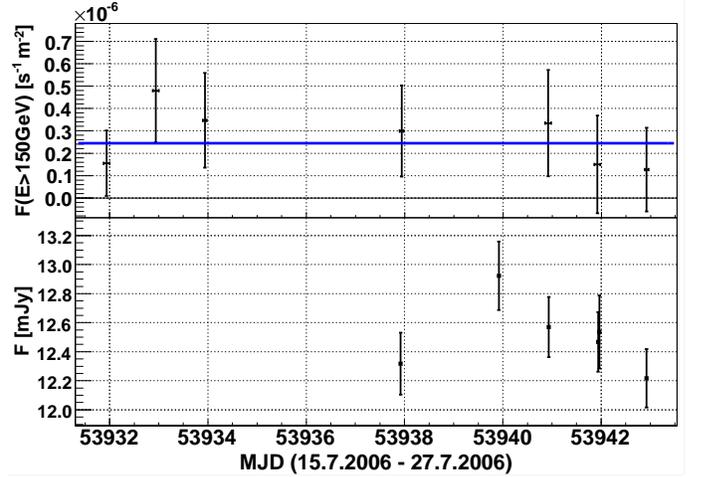}}
\caption{Night-by-night light curve for the integral flux above
150\,GeV in ${\rm ph\,s^{-1}m^{-2}}$ (upper part) and the flux in the
optical R-band (lower part) between July 15${\rm ^{th}}$ and 
27${\rm^{th}}$ 2006. For the first nights of the MAGIC observations, no
data by the optical telescope KVA were obtained. \label{lc}} 

\end{center} 
\end{figure}

\begin{table}
\centering
\caption{Date, start time, duration, and flux above 150\,GeV for
the MAGIC observations carried out between July 15${\rm ^{th}}$ and
27${\rm ^{th}}$ 2006. For one day, the flux level could not be
determined because no signal for energies above 150\,GeV was detectable
in this data set.}
\label{tablelc}
\begin{tabular}{l c}
\hline\hline
Observation  [UTC]    & F (\textgreater 150\,GeV) \\
Start Time (Duration) & [${\rm ph\,s^{-1}m^{-2}}$]\\
\hline
15.7.2006 21:45 (1.4\,h)  & $(1.55\pm1.46)\cdot10^{-7}$\\
16.7.2006 21:45 (1.37\,h) & $(4.79\pm2.31)\cdot10^{-7}$\\
17.7.2006 22:15 (0.85\,h) & $(3.47\pm2.14)\cdot10^{-7}$\\
21.7.2006 22:13 (0.62\,h) & $(2.99\pm2.04)\cdot10^{-7}$\\
23.7.2006 21:38 (0.88\,h) & -\\
24.7.2006 21:50 (1.09\,h) & $(3.34\pm2.38)\cdot10^{-7}$\\
25.7.2006 21:38 (1.31\,h) & $(1.50\pm2.18)\cdot10^{-7}$\\
26.7.2006 21:37 (0.92\,h) & $(1.27\pm1.87)\cdot10^{-7}$\\
\hline
\end{tabular}
\end{table}

To check whether a flare occurred during the eight nights of
observation, the flux above 150\,GeV was calculated on a night-by-night
basis (see Table~\ref{tablelc}). The corresponding light curve is shown
in the upper part of  Fig.~\ref{lc}. Since the data sets for single
days are of durations shorter than one and a half hours
(0.62\,h\,-\,1.4\,h), the data points represent only signals of a
significance level between 0.7\,$\sigma$ and 2.1\,$\sigma$.
Consequently, no strong conclusions about the night-to-night
variability in the flux can be drawn. Within the errors, the
measurement is consistent with a constant flux (fit probability 82\%).

Contemproraneously with the MAGIC observations, the optical telescope
KVA acquired data in the R-band. For the first nights, no data was
available. The flux for additional nights is shown in the lower part of
Fig.~\ref{lc}.

\section{Conclusion and Outlook}

PG\,1553+113 was observed in July 2006 as part of a MWL campaign with
the MAGIC telescope. After correcting the data for the effect of
calima, the analysis detected a gamma-ray signal of 5.0 standard
deviations. Within the statistical errors, the differential energy
spectrum is compatible with those derived by previous measurements
including the one observed by H.E.S.S.\ in this campaign. The
inter-night light curve shows no significant variability. 

The measured flux and reconstructed spectrum will be used in studies of
the spectral energy distribution, which will include other data
acquired during the MWL campaign \citep{mwl}.

\section*{Acknowledgements}
We would like to thank the IAC for the excellent working conditions at
the Observatorio del Roque de los Muchachos in La Palma. The support of
the German BMBF and MPG, the Italian INFN and the Spanish CICYT is
gratefully acknowledged. This work was also supported by ETH research
grant TH-34/04-3 and by Polish grant MNiI 1P03D01028.

\bibliographystyle{aa}
\bibliography{9048}

\end{document}